\documentclass[twocolumn,showpacs,aps,prl]{revtex4}

\newcommand{\ii}{\mathrm i}

\usepackage{epsf}
\usepackage{amsmath}
\usepackage{amsfonts}
\usepackage{amssymb}
\usepackage{bm}
\usepackage{bbm}
\usepackage{nicefrac}
\usepackage{xcolor}
\usepackage{pifont}
\usepackage{graphicx}
\usepackage{enumerate}
\usepackage{dcolumn}
\usepackage{maybemath}

\definecolor{darkblue}{rgb}{0,0.33,0.65}
\definecolor{darkgreen}{rgb}{0,0.65,0.33}
\definecolor{darkpurple}{rgb}{0.23,0.23,0.0}
\definecolor{darkred}{rgb}{0.33,0.0,0.0}

\begin{document}

\newcommand{\addrMST}{
Department of Physics, Missouri University of Science and Technology,
Rolla, Missouri, MO65409-0640, USA}

\title{From Dirac theories in curved space-times to a\\
variation of Dirac's large-number hypothesis}

\author{U.~D.~Jentschura}

\affiliation{\addrMST}

\begin{abstract}
An overview is given of recent developments in the field 
of Dirac equations generalized to curved space-times.
An illustrative discussion is provided. 
We conclude with a variation of Dirac's large-number 
hypothesis which relates a number of physical quantities
on cosmological scales.
\end{abstract}


\maketitle

Some physicists, even today, may think that it is impossible to connect
relativistic quantum mechanics and general
relativity~\cite{Ei1915,Hi1915,Ei1916}.
This impression, however, is false, as demonstrated, 
among many other recent works, in an article published 
not long ago in this journal~\cite{GoNe2014}.
In order to understand the subtlety, let us remember the meaning of
first~\cite{CTDiLa1978vol1,CTDiLa1978vol2}, second~\cite{ItZu1980,AbHa1986},
and third quantization (see Refs.~\cite{PiMo2000,Ki2014,Bl2014}).  In first
quantization, what was previously a well-defined particle trajectory now
becomes smeared and defines a probability density of the quantum mechanical
particle. In second quantization, what was previously a well-defined functional
value of a physical field at a given space-time point now becomes a field
operator, which is subject to quantum fluctuations.  E.g., quantum fluctuations
about the flat space-time metric can be expressed in terms of the graviton
field operator, for which an explicit expression can be found in Eq.~(5) of
Ref.~\cite{AbHa1986}.

In third quantization, what was previously a well-defined space-time point now
becomes an operator, defining a conceivably noncommutative entity, which
describes space-time quantization.  E.g., the emergence of the 
noncommutative Moyal product in
quantized string theory is discussed in a particularly clear exposition in
Ref.~\cite{Bl2014}, in the derivation leading to Eq.~(32) therein.
Speculations about conceivable effects, caused by the noncommutativity of
space-time, in hydrogen and other high-precision experiments, have been
recorded in the literature~\cite{ChSJTu2001} (these rely on noncommuting
coordinates which define a Moyal--Weyl plane). It is instructive 
to realize that relativistic quantum mechanics can easily be combined
with general relativity if the latter is formulated classically,
while the former relies on wave functions expressed in terms 
of the space-time coordinates.

We thus observe that space-time quantization is not necessary, a priori,
in order to combine general relativity and quantum 
mechanics.  Indeed, before we could ever
conceive to observe effects due to space-time quantization,
we should first consider the leading-order coupling of the
Dirac particle to a curved space-time.  Space-time curvature is visible on the
classical level, and can be treated on the classical level~\cite{Ei1916}.
Deviations from perfect Lorentz symmetry caused by effects other than
space-time curvature may result in conceivable anisotropies of space-time; they
have been investigated by Kostelecky {\em et al.} in a series of papers (see
Refs.~\cite{CoKo1997,DiKoMe2009,KoMe2012} and references therein) and would be
visible in tiny deviations of the dispersion relations of the relativistic
particles from the predictions of Dirac theory.  Measurements on
neutrinos~\cite{KoMe2012} can be used in order to constrain the
Lorentz-violating parameters.
Other recent studies concern the modification of 
gravitational effects under slight global violations of Lorentz
symmetry~\cite{KoTa2011}.

Brill and Wheeler~\cite{BrWh1957} were among the first to study the
gravitational interactions of Dirac particles, and they put a special emphasis
on neutrinos.  The motivation can easily be guessed: Neutrinos, at the time,
were thought to be massless and transform according to the fundamental
$(\tfrac12,0)$ and $(0,\tfrac12)$ representations of the Lorentz group.  Yet,
their interactions have a profound impact on cosmology~\cite{Wh1960}.  Being
(almost) massless, their gravitational interactions can, in principle, only be
formulated on a fully relativistic footing. This situation illustrates a
pertinent dilemma: namely, the gravitational potential, unlike the the Coulomb
potential, cannot simply be inserted into the Dirac Hamiltonian based on the
correspondence principle of classical and quantum
physics~\cite{CTDiLa1978vol1,CTDiLa1978vol2,SoMuGr1977,Pa1980,GiLi2013}.

For illustration, the free Dirac Hamiltonian is 
$\vec\alpha \cdot \vec p + \beta \, m$, the 
Dirac--Coulomb  Hamiltonian is 
$\vec\alpha \cdot \vec p + \beta \, m - Z\alpha/r$, 
but the Dirac--Schwarzschild~\cite{DoHo1986,Ob2001,SiTe2005,JeNo2013pra} 
Hamiltonian is {\em not} simply given as 
$\vec\alpha \cdot \vec p + \beta \, m - G \, m \, M/r$.
Here, $Z$ is the nuclear charge number, $\alpha$ is the 
fine-structure constant, $G$ is Newton's gravitational constant,
$m$ and $M$ are the masses of the two involved particles, and 
$r$ is the distance from the gravitational center.
We use the Dirac $\vec\alpha$ and $\beta$ matrices in the 
standard representation~\cite{JeNo2013pra}.
Natural units with $\hbar = c = \epsilon_0 = 1$ are employed.

Key to the calculation of the gravitational coupling of Dirac 
particles is the observation that the Dirac--Clifford algebra
needs to be augmented to include the local character of the 
space-time metric,
\begin{equation}
\label{flatcurved}
\{ \gamma^\mu, \gamma^\nu \} = 2 \, g^{\mu\nu} \to 
\{ \gamma^\mu(x), \gamma^\nu(x) \} = 2 \, g^{\mu\nu}(x)  \,,
\end{equation}
and yet, the local Dirac matrices $\gamma^\mu(x)$ have 
to be in full consistency with the requirement that the 
covariant derivative of the metric vanishes (axiom of local Lorentz 
frames in general relativity). Curved and flat space-time 
matrices are denoted with and without the argument~$x$,
respectively.
This leads to the following ansatz for the covariant derivative of a 
Dirac spinor,
\begin{equation}
\partial_\mu \psi(x) \to \nabla_\mu \psi(x) = 
\left[ \partial_\mu -\Gamma_\mu(x) \right] \, \psi(x)  \,,
\end{equation}
and the spin-connection matrices $\Gamma^\mu(x)$ are expressed
in terms of the $\gamma^\mu(x)$ as follows~\cite{BrWh1957,Je2013pra},
\begin{equation}
\Gamma_\rho = - \frac{\ii}{4} \, g_{\mu\alpha}(x) \,
\left( \frac{\partial {b_\nu}^\beta(x)}{\partial x^\rho} \,
{a^\alpha}_\beta(x) - {\Gamma^\alpha}_{\nu \rho} \right) \,
\sigma^{\mu\nu}(x),
\end{equation}
where $\sigma^{\mu\nu}(x) = 
\frac{\ii}{2} [\gamma^\mu(x), \gamma^\nu(x)]$ is a local 
spin matrix and the $a$ and $b$ coefficients belong to the 
square root of the metric,
$\gamma_\rho(x) = {b_\rho}^\alpha(x) \, \gamma_\alpha$,
and
$\gamma^\alpha(x) = {a^\alpha}_\rho(x) \, \gamma^\rho$,
while the Christoffel symbols are 
${\Gamma^\alpha}_{\nu \rho} \equiv {\Gamma^\alpha}_{\nu \rho}(x)$.
This formalism was later shown to be compatible with the 
exponentiation of the local generators~\cite{Bo1975prd,SoMuGr1977,GiLi2013} 
of the Lorentz 
group in the spinor representation, as explained for flat space-time 
in detail in Ref.~\cite{ItZu1980}.
Contrary to wide-spread belief,
it is indeed possible to combine relativistic quantum 
mechanics and general relativity, at least on the level 
of quantum mechanics and quantum fields, 
in curved space-times, which are defined 
by a nontrivial (but classical, non-quantized) 
metric $g^{\mu\nu}(x)$. This formalism
is free from ambiguities and has been used in order
to evaluate perturbative corrections to the hydrogen spectrum
in curved space-times with a nontrivial metric~\cite{Pa1980prd,PaPi1982}.

Recent developments in this field include a series of 
papers~\cite{HeNi1990,Ob2001,SiTe2005,SiTe2007,ObSiTe2009,ObSiTe2011,
Je2013pra,JeNo2013pra,JeNo2014jpa,Je2014pra},
where the gravitationally coupled Dirac particle is formulated first in a fully
relativistic setting, and then, a generalized Foldy---Wouthuysen transformation
is applied in order to  isolate the nonrelativistic limit, plus correction
terms.  For the photon emission in curved space-time, we refer to the result
for the vector transition current in Ref.~\cite{JeNo2013pra}.
A certain pitfall, hard to spot at first glance because a deceptively
elegant Eriksen--Kolsrud~\cite{ErKo1958} variant of the standard Foldy--Wouthuysen
transformation exists, leads to the rather
unfortunate appearance of spurious, parity-violating terms.
In Ref.~\cite{JeNo2014jpa}, the conclusion 
has been reached that, in a typical, nontrivial 
space-time geometry, the parity-breaking terms in the in the 
``chiral'' modification~\cite{ErKo1958}
of the Foldy---Wouthuysen transformation change the interpretation
of the wave functions and of the transformed operators to the 
extent that the transformed entities cannot be associated any more
with their ``usual'' physical interpretation after the transformation,
questioning the usefulness of the ``chiral'' transformation
(for more details, see Sec.~4 of Ref.~\cite{JeNo2014jpa}).
We here take the opportunity to discuss the Foldy--Wouthuysen
transformation of gravitationally coupled Dirac particles in some further
detail.  Of particular importance is the 
Dirac--Schwarzschild Hamiltonian~\cite{Ob2001,SiTe2005,JeNo2013pra},
which describes the most basic central-field problem in gravitation and
probably replaces the Dirac--Coulomb Hamiltonian~\cite{SwDr1991a,SwDr1991b,SwDr1991c} 
as the paradigmatic bound-state problem in gravitation.

\begin{figure}[t!]
\begin{center}
\begin{minipage}{1.0\linewidth}
\includegraphics[width=0.81\linewidth]{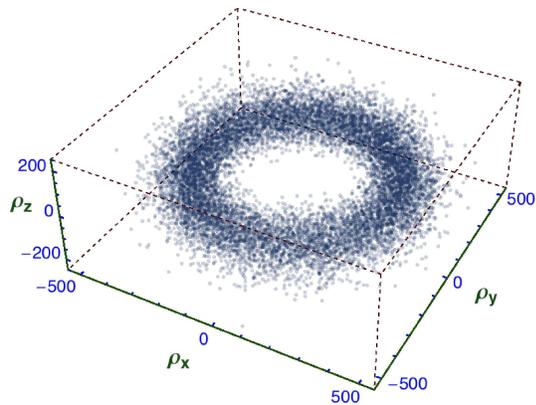}
\caption{\label{fig1} 
Probability density of a neutron in a quantum Rydberg state
with $n=19$ and maximum orbital angular momentum 
quantum number $\ell=m=18$, gravitationally bound to a 
(tiny) black hole located at the origin.
The scaled coordinates $\vec \rho$ is 
defined in the text immediately following Eq.~\eqref{HS_grav}.
For a neutron bound to a black hole of mass
equal to $10^{-15}$ times the mass of the Earth,
the gravitational fine-structure constant has the numerical 
$\alpha_G = 0.0211$.  The points are distributed 
with a probability density proportional to $|\psi|^2$,
i.e., proportional to the probability of finding the
electron at a particular coordinate in the immediate vicinity 
of the plotted coordinate. The circular Rydberg 
state approximates a classical circular trajectory with 
$\sqrt{ \langle \vec\rho^2 \rangle} = 375.22$.}
\end{minipage}
\end{center}
\end{figure}

Inherently, the Foldy--Wouthuysen program is perturbative;
for nontrivial geometries, one tries to disentangle the particles
and the antiparticle degrees of freedom which are otherwise 
{\em simultaneously} described by the Dirac theory~\cite{Je2013pra}.
A Hamiltonian, which preserves the full parity symmetry 
of the Schwarzschild geometry, and which allows for a clear 
identification of the relativistic corrections terms, 
has recently been presented in Ref.~\cite{JeNo2013pra}. It reads
\begin{align}
\label{HFW}
& H_{\rm FW} =
 \beta \, \left( m + \frac{\vec p^{\,2}}{2 m} - 
\frac{\vec p^{\,4}}{8 m^3} \right) 
- \beta \, \frac{m \, r_s}{2 \, r} 
\\[1.0ex]
& \; + \beta \, \left( 
- \frac{3 r_s}{8 m} \, 
\left\{ \vec p^{\,2}, \frac{1}{r} \right\}
+ \frac{3 \pi r_s}{4 m} \delta^{(3)}(\vec r) \,
+ \frac{3 r_s}{8 m} \, \frac{\vec\Sigma \cdot \vec L}{r^3} \right) \,.
\nonumber
\end{align}
Here, the Schwarzschild radius is given as
$r_s = 2 \, G \, M$. This Hamiltonian 
is equivalent to the one presented in Eq.~(26) 
of the recent paper published in this journal~\cite{GoNe2014},
provided the $V$ and $W$ functions are expanded
to first order in the gravitational constant $G$ (first order in the 
Schwarzschild radius $r_S$). 
The subject is not without pitfalls:
We note that Donoghue and Holstein~\cite{DoHo1986}, two authors 
otherwise known for their 
extremely careful analysis of a number of physical problems, obtained 
spurious parity-violating terms after the Foldy--Wouthuysen 
transformation
[see the text surrounding Eq.~(46) of Ref.~\cite{DoHo1986}];
the community seems to have converged on the 
result that these terms actually are absent.
The manifestly and fully parity-conserving Hermitian form~\eqref{HFW}
allows for an immediate answer to a pressing question:
If the electrodynamic interaction were absent,
would the quantum mechanical Schr\"{o}dinger problem 
of proton and electron still have a solution and feature bound states?
The answer is yes. The leading nonrelativistic terms from Eq.~\eqref{HFW}
exactly have the structure expected from a naive insertion of the gravitational
potential into the Schr\"{o}dinger Hamiltonian, namely
\begin{align}
\label{HS_grav}
H_S =& \; 
\frac{\vec p^{\,2}}{2 m} - \frac{m \, r_s}{2 \, r}
= \alpha_G^2 m_e  c^2 \, \left( -\frac12 \, \vec\nabla^2_\rho - 
\frac{1}{\rho} \right) \,,
\end{align}
where the gravitational fine-structure
constant is $\alpha_G = G \, m_e \, m_p/(\hbar c)$
and $\rho = G m_e m_p r/\hbar^2$.
The relativistic corrections 
are described by the expectation value of the remaining terms
in Eq.~\eqref{HFW} on Schr\"{o}dinger--Pauli wave functions and lead to the 
gravitational fine-structure formula~\cite{Je2014pra}
\begin{equation}
\label{mainres}
E_{n \varkappa} = 
-\frac{\alpha_G^2 m_e c^2}{2 n^2} +
\frac{\alpha^4_G m_e c^2}{n^3} \, \left( \frac{15}{8 n} -
\frac{14 \, \varkappa + 3}{2 \,|\varkappa| \, (2 \varkappa + 1)} \right)\,,
\end{equation}
where $n$ is the principal quantum number,
$\varkappa$ is the Dirac angular quantum number,
which summarizes both the orbital angular momentum 
$\ell$ as well as the total angular momentum $j$ of the 
bound particle into a single integer, 
according to the formula
$\varkappa = (-1)^{j+\ell+1/2} \, \left(j + \frac12 \right)$.
Initial steps toward the calculation of the 
fine-structure formula had been taken in Refs.~\cite{GaPoCh1983,VrEtAl2013};
the formula~\eqref{mainres} is in agreement with 
the form of the gravitational Hamiltonian given in Eq.~(26)
of Ref.~\cite{GoNe2014} and with Refs.~\cite{JeNo2013pra,JeNo2014jpa}. 
The energy levels~\eqref{mainres}
lift the $(n,j)$ degeneracy known from the Dirac--Coulomb problem.
For electron-proton interactions,
the gravitational fine-structure constant
$\alpha_G \approx 3.2164 \times 10^{-42} $
is a lot smaller than the electromagnetic 
fine-structure constant $\alpha_{\rm QED} \approx 1/137.036$,
but $\alpha_G$ can be a lot larger for particles
bound to black holes even if the ``tiny'' black holes
are lighter than the Earth.
A ``scatter plot'' of a gravitational bound state (circular Rydberg state) 
which illustrates the quantum-classical correspondence
is given in Fig.~\ref{fig1}.

Dirac's large-number hypothesis~\cite{Di1938LNH} is based on the observation 
that the ratio 
\begin{equation}
\frac{\alpha_{\rm QED}}{\alpha_G} =
\frac{e^2}{4 \pi \epsilon_0 G \, m_e \, m_p} \approx 2.3 \times 10^{39}  
\approx \frac{c \, T}{r_e}
\end{equation}
is approximately equal to the ratio of the age of the 
Universe, $T$, to the time it takes light to travel 
a distance equal to the classical electron radius,
$r_e/c$, where $r_e$ is the classical electron radius,
$r_e = \alpha \hbar/(m_e c)$.
Eddington~\cite{Ed1931} observed that the ratio of 
$\alpha_{\rm QED}$ to the gravitational fine-structure 
constant $\alpha^{(ee)}_G$ for two gravitating 
electrons is approximately equal to the square root 
of the number $N$ of charged particles in the Universe,
$\alpha_{\rm QED}/\alpha^{(ee)}_G =
e^2/(4 \pi \epsilon_0 G \, m^2_e) \approx 4.2 \times 10^{42}  
\approx \sqrt{ \, N \,} $.
One may also curiously observe, as a variation 
of the other observations recorded in the literature, that 
\begin{equation}
\label{spec}
\frac{1}{\sqrt{\alpha_G}} \,
\exp\left( -\sqrt{\frac{m_p}{m_e}} \right) 
\approx 137 \approx \frac{1}{\alpha_{\rm QED}} \,.
\end{equation}
With the most recent {\rm CODATA} value~\cite{MoTaNe2012}
for $G$ plugged into the left-hand side of Eq.~\eqref{spec}, the 
left-hand side and right-hand sides of Eq.~\eqref{spec} are 
in agreement to better than half a permille.
It is unclear at present if the emergence 
of the famous factor ``$137$'' from an equation 
whose only adjustable prefactor is the number one
[see Eq.~\eqref{spec}] constitutes a pure accident, or if
there is a deeper physical interpretation available.
Variants of unified electromagnetic and 
gravitational theories, with Kaluza--Klein compactification,
may lead to massless, or mass-protected spinors coupled
to the gauge fields~\cite{MBNi2006}.
Possible connections of gravitational and electromagnetic 
interactions have intrigued physicists ever since
Gertsenshtein~\cite{Ge1962}, as well as 
Zeldovich and Novikov~\cite{ZeNo1983}, discovered the 
possibility of graviton-photon conversion.

Work on this project has  been supported by the National Science Foundation 
(Grant PHY--1403973) and by the National Institute of Standards and Technology
(precision measurement grant).

\end{document}